\documentclass[aps,showpacs,a4paper,superscriptaddress]{revtex4}
\usepackage{hyperref}
\usepackage{amsmath}
\usepackage{amssymb}
\usepackage{graphicx}
\usepackage{varwidth,array}

\newcommand{\ep}{\varepsilon}  

\def\vec#1{\boldsymbol{#1}}
\usepackage{ulem} 
\usepackage{color}

\begin{document}
\title{Principal spectra describing magnetooptic permittivity tensor in cubic crystals}
\author{Jana Hamrlov\'a}
\affiliation{Nanotechnology Centre, VSB -- Technical University of Ostrava, 17. listopadu 15, 708 33 Ostrava, Czech Republic}
\affiliation{IT4Innovations Centre, VSB -- Technical University of Ostrava, 17. listopadu 15, 708 33 Ostrava, Czech Republic}
\author{Dominik Legut}
\affiliation{IT4Innovations Centre, VSB -- Technical University of Ostrava, 17. listopadu 15, 708 33 Ostrava, Czech Republic}
\author{Martin Veis}
\affiliation{Faculty of Mathematics and Physics, Charles University, Ke Karlovu 3, 121 16 Prague, Czech Republic}
\author{Jarom\'{\i}r Pi\v{s}tora}
\affiliation{Nanotechnology Centre, VSB -- Technical University of Ostrava, 17. listopadu 15, 708 33 Ostrava, Czech Republic}
\author{Jaroslav Hamrle}
\affiliation{IT4Innovations Centre, VSB -- Technical University of Ostrava, 17. listopadu 15, 708 33 Ostrava, Czech Republic}
\affiliation{Faculty of Mathematics and Physics, Charles University, Ke Karlovu 3, 121 16 Prague, Czech Republic}
\affiliation{Department of Physics, VSB -- Technical University of Ostrava, 17. listopadu 15, 708 33 Ostrava, Czech Republic}
\date{\today}

\begin{abstract}
   We provide unified phenomenological description of magnetooptic effects being linear and quadratic in magnetization. The description is based on few principal spectra, 
   describing elements of permittivity tensor up to the second order in magnetization. Each permittivity tensor element for any magnetization direction and any sample surface orientation is simply determined by weighted summation of the principal spectra, where weights are given by crystallographic and magnetization orientations. The number of principal spectra depends on the symmetry of the crystal. In cubic crystals owning point symmetry we need only four principal spectra. Here, the principal spectra are expressed by {\it ab-initio} calculations for bcc Fe, fcc Co and fcc Ni in optical range as well as in hard and soft x-ray energy range, {\it i.e.} at the 2$p$- and 3$p$-edges. We also express principal spectra analytically using modified Kubo formula. 
\end{abstract}
\pacs{42.50.Ct, 31.15.A-, 78.20.Ls, 78.70.Dm, 78.40.Kc}
\maketitle

There is a vast number of physical phenomena proportional to quadratic form of magnetization. In case of dc transport phenomena, the most well-known examples are anisotropy magnetoresistance (AMR) \cite{mallinson,doring1938} or longitudinal Hall effect \cite{grebner60}. 
Within the magnetooptic community, the field of magnetooptic effects quadratic in magnetization is complicated by incredible number of nomenclature, being called Cotton-Mouton effect, Voigt effect, quadratic magnetooptic Kerr effect (QMOKE) \cite{ham07quad}, magnetic linear birefringence, X-ray magnetic linear dichroism (XMLD) \cite{mertins01,valencia10}, magnetic double refraction, magnetooptic orientation effect, magnetooptic anisotropy, Hubert-Sch\"{a}fer effect or magnetorefractive effect \cite{hrabovsky2009,kravets2002}. The nomenclature is not strictly defined, however it refers either to type of samples (liquid, gas, solid state) or it refers to experimental configurations of the setup (namely detecting change of light intensity or detecting change of polarization state upon variation of magnetization direction). Although those effects are usually not considered as single phenomena, they originate from equal parts of permittivity tensors (i.e.\ from equal symmetry breaking). 
Notice, that recently new types of quadratic-in-magnetization effects arose, for example anisotropic magneto-thermopower \cite{avery2012,schmid2013} in spin-caloritronics. Within some generalization, one can expect that any magneto-transport linear in magnetization will have its quadratic-in-magnetization counterpart.

The first observation of magnetooptic effects quadratic in magnetization dates back more than century ago in works of Kerr \cite{kerr1901}, Majorana \cite{majorana1902} and Cotton and Moutton \cite{cotton1905}, where magnetic birefringence was observed in liquids and colloids. Later, those quadratic magnetooptic response was observed in gases, solid state materials and obviously also in ferromagnetic materials. See large reviews of Smolenskii \textit{et al} \cite{pisarev1975} and Ferr\'e, Gehring \cite{ferre1984} from 80's. The anisotropy of magnetooptic effects (i.e. dependence of QMOKE on crystal and field orientations) was investigated for various systems. However, the investigation were mostly done for a single photon energy, for example Fe \cite{pos02} or antiferromagnetic RbMnF$_3$ \cite{ferre1983}. Buchmaier \textit{et al} have even determined values of principal spectra (namely values of $2G_{44}$ and $\Delta G$), however also only for a single photon energy \cite{buchmeier2002}. The spectral 
investigations were rare, for example spectroscopy of QMOKE for different orientations of Ni surfaces by Krinchik \cite{krinchik1971} and Parker \cite{parker1972}. 

The evolution of quadratic magnetooptic effects in X-ray photon energy range was rather independent on those in extended visible range. The x-ray magnetic linear dichroism were theoretically predicted by Thole and van~der~Laan \cite{Thole85,laan86}. Due to experimental restrictions, the quadratic-in-magnetization effects in X-ray range were measured in absorption, {\it i.e.} dichroic effects \cite{stohr99,scholl2000} and later in polarization, Voigt effect \cite{mertins01}. The XMLD was analytically predicted for [100] and [111] magnetization axis by Kunes \textit{et al} \cite{Kunes03}. They showed that for the correct shape of the XMLD spectra at L$_{2,3}$ edges the exchange splitting of the core levels is necessary \cite{kun04}. Later, with direct comparison with the experiment the angular dependence of the XMLD of bcc Fe shown that the role of the valence spin-orbit interaction is much weaker \cite{Nol10}. The simple model of Kunes \textit{et al} that accounts for 
the difference of the XMLD spectra for the high symmetry crystal directions, \textit{et. al} with magnetization along [001] and [111] axis showed that the origin is expressed as different weighted spin densities of $e_g$ and $t_{2g}$ orbitals \cite{kunes2003}. Shortly after, Arenholz \textit{et. al} \cite{arenholz2006,arenholz2007} have shown that in case of cubic crystals having point symmetry, XMLD along any quantization axis can be expressed as weighted sum of two fundamental spectra. 

This approach was extended to tetragonal crystal field by Laan \textit{et al}, extending number of fundamental spectra to four, demonstrated on CoFe$_2$O$_4$ \cite{laan2008}.
Krug \textit{et al} \cite{krug2008} have generalize this approach from XMLD fundamental spectra to fundamental spectra of scattering form-factor $\mathcal{F}$, where imaginary part of $\mathcal{F}$ corresponds to light absorption. Later, Havekort \textit{et al} have expressed symmetry arguments also for other crystal symmetries \cite{havekort2010}. Recently, the quadratic effects in magnetization like Voigt in reflection or XMLD were detected and theoretically explained also in the soft x-ray energy range, \textit{i. e.} at the 3$p$ edges of transition metals \cite{valencia10, leg14, tesch2014, mertins2014}.


From phenomenological point of view, all magnetooptic effects are described by conductivity $\sigma_{ij}$, permittivity $\ep_{ij}$ or scattering $F_{ij}$ tensors, being simply convertible between each other. In following, we use tensor of permittivity, being mostly employed in optical community. From quantum-mechanical point of view, the fundamental part of the permittivity tensor is optical absorption, related to optical dispersion by Kramers-Kronig relations \cite{rathgen2004}. The relations of magnetooptic response (or in general to any magneto-transport) to crystallographic orientation is given by symmetry arguments, predicting dependence of the permittivity tensor elements on magnetization direction for effects being independent, linear, quadratic, cubic etc in magnetization. In case of magnetooptic effects quadratic-in-magnetization, the form of such tensors can be found in works of Birss \cite{birss}, Nye \cite{nye} and Bhagavantam \cite{bhagavantam} and Visnovsky \cite{visnovsky1986}, whose 
notation is used in this article. Symmetry arguments predict, that the permittivity tensor is given by a weighted sum of constituent single dielectric tensor elements, where their weights are given solely by direction of magnetization. 

In this article, we express principal spectra, \textit{i.e.} the non-magnetic, linear and quadratic contributions in magnetization for permittivity tensors, from which the complete spectra of permittivity tensor can be obtained for any magnetization direction. Then, we express principal spectra calculated from first principles of cubic material bcc Fe, fcc Co and fcc Ni for three spectral ranges, extended visible range, soft and hard x-rays (at the 2$p$-edges and 3$p$-edges, respectively). Also, we express principal spectra directly using modified Kubo formula. Finally, we briefly express relations between signal measured by various magnetooptical techniques and off-diagonal permittivity elements.

\section{Permittivity tensor in cubic crystals}

The optical or conductivity properties of generally anisotropic material are described by (generally complex) permittivity $\ep_{ij}$ or conductivity $\sigma_{ij}$ tensor, being mutually related by (in SI units) $\ep_{ij}=i \sigma_{ij}/\omega +\delta_{ij}$, where $\delta_{ij}$ being Kronecker delta and $\omega$ light frequency (i.e.\ $\omega=0$ corresponds to d.c.\ current). Onsager relations require antisymmetry in magnetization direction, $\ep_{ij}(\vec{M})=\ep_{ji}(-\vec{M})$. As various magnetization direction causes small perturbation to the crystal optical properties, $\ep_{ij}$, they can be expressed as Taylor series of permittivity tensor in direction of $\vec{M}$ \cite{visnovsky1986}
\begin{equation}
\label{eq:eptaylor}
\varepsilon_{ij} = \varepsilon_{ij}^{(0)}+K_{ijk}M_k+G_{ijkl}M_kM_l+\ldots  = \varepsilon_{ij}^{(0)}+\varepsilon_{ij}^{(1)}+\varepsilon_{ij}^{(2)} +\ldots
\end{equation}
where superscripts denote zeroth, first, second etc.\ order in magnetization. Optical permittivity independent on magnetization direction is $\ep_{ij}^{(0)}$ and in case of cubic crystal, it has unitary form $\ep_{ij}^{(0)}=\ep^{(0)}\delta_{ij}$. 
The term linear in magnetization $\varepsilon_{ij}^{(1)}=K_{ijk}M_k$ is described by third-order tensor $K_{ijk}$, where Onsager relations assures $\ep^{(1)}_{ij}=-\ep^{(1)}_{ji}$. In case of cubic crystals, it has simple form $K_{ijk} = \epsilon_{ijk} K$, where $\epsilon_{ijk}$ is Levi-Civita symbol and $K$ is linear magnetooptic element.

The term quadratic in magnetization $\varepsilon_{ij}^{(2)}=G_{ijkl}M_kM_l$ is described by fourth order tensor $G_{ijkl}$, with Onsager relations assuring $\ep^{(2)}_{ij}=\ep^{(2)}_{ji}$. In case of cubic crystal with $\hat{x}\parallel[100]$, $\hat{y}\parallel[010]$, $\hat{z}\parallel[001]$, the $G_{ijkl}$ tensor can be written in compact form as \cite{birss, nye, bhagavantam, visnovsky1986}
\begin{equation}
\label{eq:Gcubic}
\left[ 
\begin{array}{c}
\varepsilon_{xx}^{(2)}\\
\varepsilon_{yy}^{(2)}\\
\varepsilon_{zz}^{(2)}\\
\varepsilon_{yz}^{(2)}\\
\varepsilon_{zx}^{(2)}\\
\varepsilon_{xy}^{(2)}
\end{array}
\right] = \left[ 
\begin{array}{cccccc}
G_{11} & G_{12} & G_{21} & 0 & 0 & 0 \\
G_{21} & G_{11} & G_{12} & 0 & 0 & 0 \\
G_{12} & G_{21} & G_{11} & 0 & 0 & 0 \\
0 & 0 & 0 & 2G_{44} & 0 & 0 \\
0 & 0 & 0 & 0 & 2G_{44} & 0 \\
0 & 0 & 0 & 0 & 0 & 2G_{44} 
\end{array}
\right] \,
\left[ 
\begin{array}{c}
M_x^2 \\
M_y^2 \\
M_z^2\\
M_yM_z\\
M_zM_x\\
M_xM_y
\end{array}
\right]
\end{equation}

In order to express $G$-elements for crystals with cubic symmetry, the elements can be rearranged defining $G_s = G_{11}-(G_{12}+G_{21})/2$ and 
$\Delta\Gamma=(G_{12}-G_{21})/2$. Note that alternative way used to express anisotropy of the permittivity  is $\Delta G=G_{11}-2G_{44}-(G_{12}+G_{21})/2$ \cite{pos02,hamrlova2013}. 
In case of cubic crystals owning point O$_h$ symmetry, $G_{12}=G_{21}$ and hence $G_s=G_{11}-G_{12}$, $\Delta \Gamma=0$ \cite{bhagavantam}. 
In case of isotropic medium, then both $\Delta G=0$, $\Delta \Gamma=0$.
Furthermore, notice, that spectra of $G_{11}$ element can not be determined (only value of $G_s$) as for the ferromagnetic materials, the magnitude of magnetization vector $|\vec{M}|$ is constant for any magnetization direction and hence $G_{11}$ becomes inseparable part of $\ep^{(0)}$ .

Hence, spectra of any permittivity element $\ep_{ij}$ (up to second order in magnetization) can be expressed as weighted sum of principal spectra, where the weights are determined by crystallographic and magnetization directions. Table~\ref{t:scans} demonstrates those weights for $\hat{x}\parallel [100]$, $\hat{y}\parallel[010]$ and $\hat{z}\parallel[001]$ oriented crystal for two magnetization scans, used later in ab-initio calculations. Namely, the angular dependence  of $\ep_{ij}$ permittivity elements is expressed for two magnetization scans, in-plane ($\theta=90^\circ$, $\varphi\in\langle 0;\frac{\pi}{2} \rangle$) and out-of-plane ($\theta\in\langle 0;\frac{\pi}{2} \rangle$, $\varphi=45^\circ$) scans as the magnetization passes through all high symmetry directions [001], [110] and [111]. Angles $\theta$, $\phi$ define magnetization direction, being $\vec{M}= [M_x, M_y, M_z] =[\sin\theta \cos\varphi, \sin\theta \sin\varphi, \cos\theta ]$.
See Ref.~\cite{hamrlova2013} providing weighting factors of principal spectra for general magnetization direction and several crystallographic orientations.


\begin{table*}
\begin{tabular}{cc}
In-plane scan: $\theta=\frac{\pi}{2}$, $\varphi\in\langle0;\frac{\pi}{2}\rangle$: & 
Out-of-plane scan: $\theta\in\langle0;\frac{\pi}{2}\rangle$, $\varphi=\frac{\pi}{4}$:
\\
\begin{tabular}{|c|c|}
 \hline
  $\ep_{xx}^{(0)}$ & $\ep^{(0)}$ 
  \\ \hline
  $\ep_{yy}^{(0)}$ & $\ep^{(0)}$ 
  \\ \hline
  $\ep_{zz}^{(0)}$ & $\ep^{(0)}$ 
  \\ \hline
  $\ep_{yz}^{(1)}$ & $K\cos\varphi$
  \\ \hline
  $\ep_{zx}^{(1)}$ & $K\sin\varphi$
  \\ \hline
  $\ep_{xy}^{(1)}$ & $0$  
  \\ \hline
\end{tabular}
\begin{tabular}{|c|c|}
 \hline
  $\ep_{xx}^{(2)}$ & $G_{11}-G_{s}\sin^2\varphi$ 
  \\ \hline
  $\ep_{yy}^{(2)}$ & $G_{11}-G_{s}\cos^2\varphi$ 
  \\ \hline
  $\ep_{zz}^{(2)}$ & $G_{11}-G_{s}$ 
  \\ \hline
  $\ep_{yz}^{(2)}$ & $0$
  \\ \hline
  $\ep_{zx}^{(2)}$ & $0$
  \\ \hline
  $\ep_{xy}^{(2)}$ & $G_{44}\sin2\varphi$  
  \\ \hline
\end{tabular}
& 
\begin{tabular}{|c|c|}
 \hline
  $\ep_{xx}^{(0)}$ & $\ep^{(0)}$ 
  \\ \hline
  $\ep_{yy}^{(0)}$ & $\ep^{(0)}$ 
  \\ \hline
  $\ep_{zz}^{(0)}$ & $\ep^{(0)}$ 
  \\ \hline
  $\ep_{yz}^{(1)}$ & $K\frac{1}{\sqrt{2}}\sin\theta$
  \\ \hline
  $\ep_{zx}^{(1)}$ & $K\frac{1}{\sqrt{2}}\sin\theta$
  \\ \hline
  $\ep_{xy}^{(1)}$ & $K\cos\theta$  
  \\ \hline
  \end{tabular}
 \begin{tabular}{|c|c|}
 \hline
  $\ep_{xx}^{(2)}$ & $G_{11}+G_{s}(\frac{1}{2}\sin^2\theta-1)$ 
  \\ \hline
  $\ep_{yy}^{(2)}$ & $G_{11}+G_{s}(\frac{1}{2}\sin^2\theta-1)$ 
  \\ \hline
  $\ep_{zz}^{(2)}$ & $G_{11}-G_{s}\sin^2\theta$ 
  \\ \hline
  $\ep_{yz}^{(2)}$ & $\frac{1}{\sqrt{2}}G_{44}\sin2\theta$
  \\ \hline
  $\ep_{zx}^{(2)}$ & $\frac{1}{\sqrt{2}}G_{44}\sin2\theta$
  \\ \hline
  $\ep_{xy}^{(2)}$ & $G_{44}\sin^2\theta$  
  \\ \hline
  \end{tabular}
\end{tabular}
\caption{\label{t:scans}%
The angular dependence of permittivity elements $\ep_{ij}=\ep^{(0)}_{ij}+\ep_{ij}^{(1)}+\ep_{ij}^{(2)}$ on magnetization direction, as predicted by symmetry arguments for cubic crystals with $O_h$ point symmetry (hence $\Delta\Gamma=0$). The superscript denotes the order of magnetization for a given permittivity element. Unlisted permittivity elements are zero.
}
\end{table*}

\section{Determination of principal spectra}

\begin{table}
\begin{tabular}{ll}
$\mathbf{M}\!\!\parallel\!\![100]$ ($\theta=\frac{\pi}{2}$, $\varphi=0$):
&
\mbox{} $G_{s}=\varepsilon_{xx}^{[100]}-\varepsilon_{yy}^{[100]}$ 
\\ \hline
$\mathbf{M}\!\!\parallel\!\![110]$ ($\theta=\frac{\pi}{2}$, $\varphi=\frac{\pi}{4}$):
&
\mbox{} $G_{s}=2(\varepsilon_{yy}^{[110]}-\varepsilon_{zz}^{[110]})$ 
\\ 
&
\mbox{} $2G_{44}=\varepsilon_{xy}^{[110]}+\varepsilon_{yx}^{[110]}$ 
\\ \hline
$\mathbf{M}\!\!\parallel\!\![111]$ ($\cos\theta=\frac{1}{\sqrt{3}}$, $\varphi=\frac{\pi}{4}$):
&
\mbox{} $2G_{44}=\frac{3}{2}(\varepsilon_{xy}^{[111]}+\varepsilon_{yx}^{[111]})$
\end{tabular}
\caption{\label{t:fund}%
The $G$-elements, as determined from tensor permittivity $\ep_{ij}$ with magnetization pointing in fundamental directions. Here, the superscript denotes magnetization direction for a given tensor.
}
\end{table}

\begin{figure}
\includegraphics[width=\textwidth]{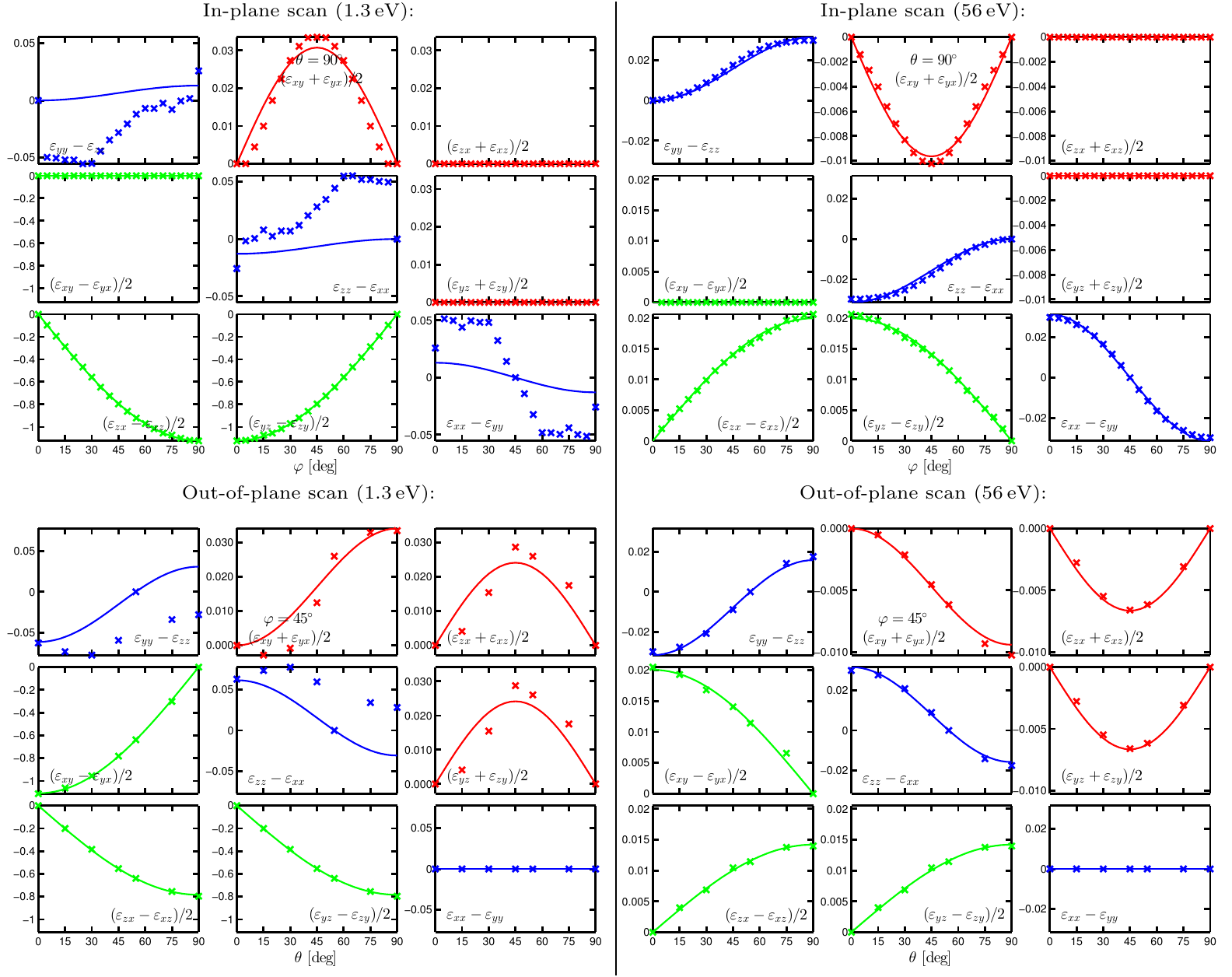}
\caption{\label{f:epsdiff}
Showcase of imaginary part of the permittivity for bcc Fe at 1.3\,eV (left columns) and 56\,eV (right columns) for differences of permittivity tensor elements on magnetization direction for (top line) in-plane scans and (bottom line) out-of-plane scans. Line is fit to dependence as predicted by symmetry arguments (Tab.~\protect\ref{t:scans}). Diagonal elements (blue) corresponds to difference of diagonal permittivity elements $\ep_{ii}-\ep_{jj}$, $i\neq j$, being proportional to $G_s=G_{11}-G_{12}$. Lower triangle (green) $(\ep_{ij}+\ep_{ji})/2$ is proportional to linear magnetooptic $K$ element. Upper triangle (red) $(\ep_{ij}-\ep_{ji})/2$ is proportional to $2G_{44}$ element.}
\end{figure}

\begin{figure*}
\includegraphics[width=\textwidth]{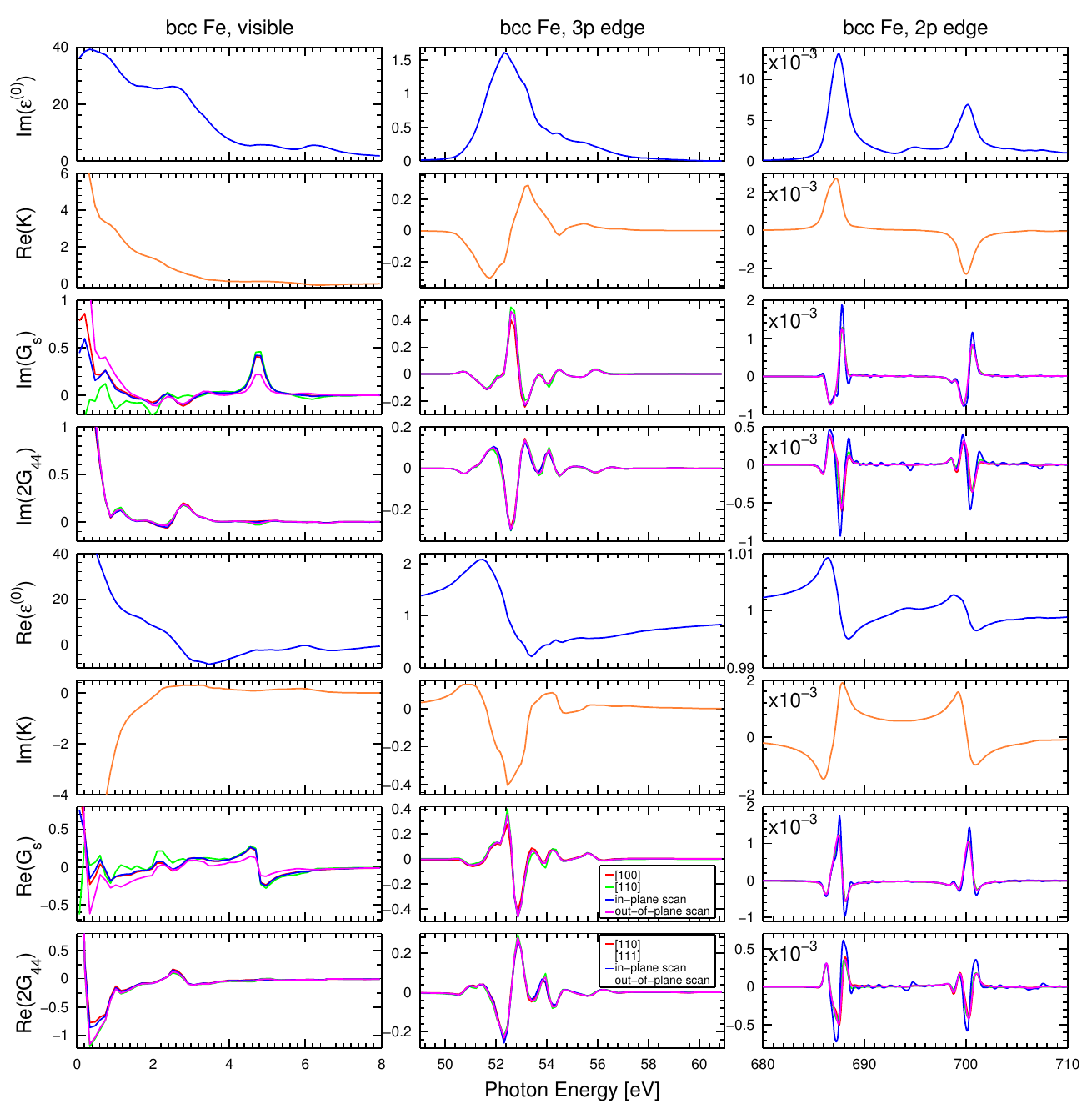}
\caption{\label{f:specFe}%
Principal spectra of permittivity elements up to the second order in magnetization ($\ep^{(0)}$, $K$, $G_s=G_{11}-G_{12}$, $2G_{44}$) for bcc Fe. Three spectral ranges are presented, extended visible range, 3$p$-edges and 2$p$-edges range. Spectra of second-order elements $G_{s}$ and $2G_{44}$ are determined by several ways, by (a) fit to dependence of $\ep_{ij}$ on magnetization direction and (b) from permittivity tensor calculated for magnetization in fundamental directions [001], [011] and [111]. The principal spectra are complex functions, top four lines show spectra parts (real or imaginary) related to the light absorption, bottom four lines show spectra parts related with light dispersion.
}
\end{figure*}

\begin{figure*}
\includegraphics[width=\textwidth]{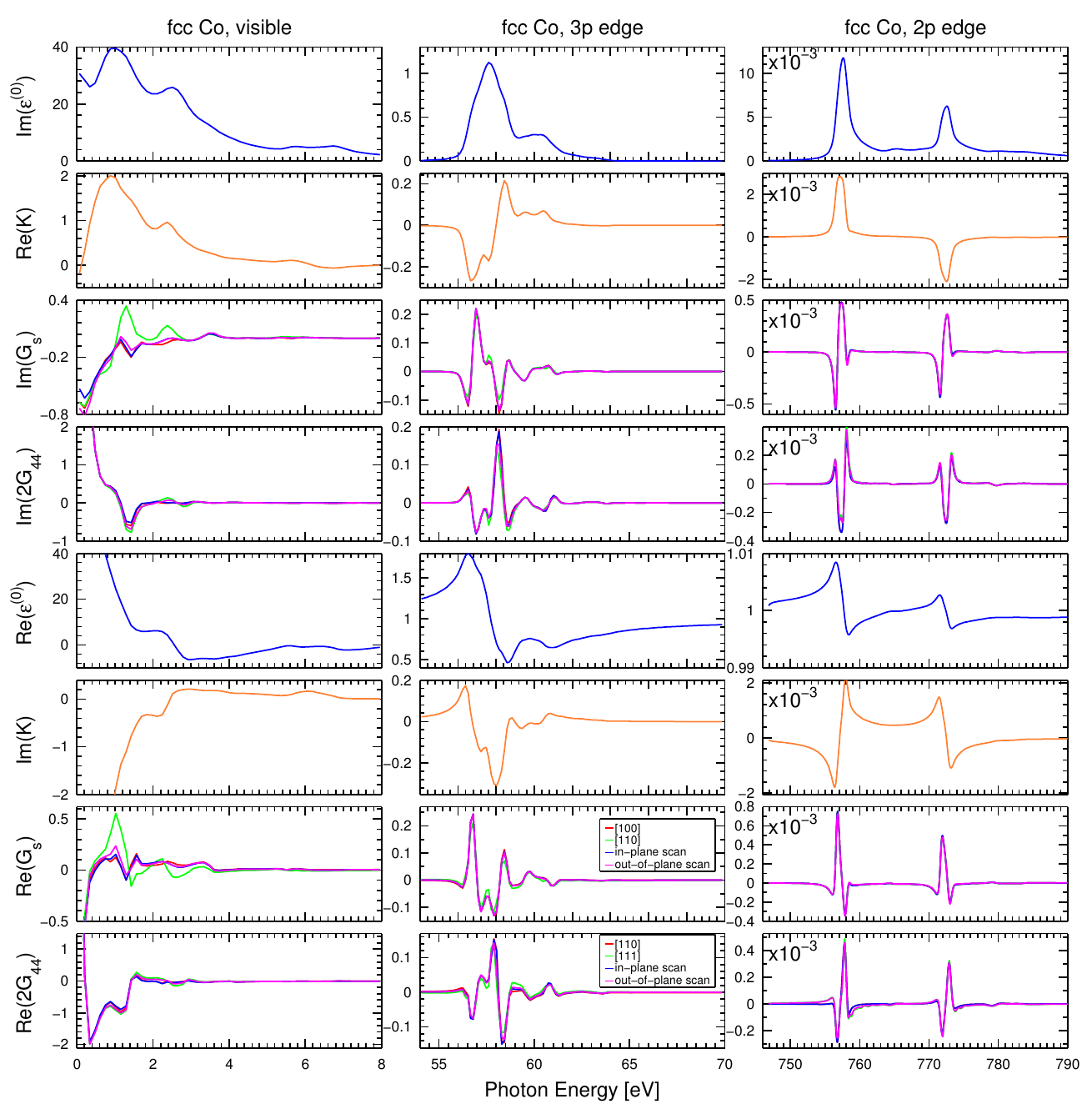}
\caption{\label{f:specCo}%
Principal spectra for fcc Co. Remaining description is the same as in Fig.~\protect\ref{f:specFe}.
}
\end{figure*}

\begin{figure*}
\includegraphics[width=\textwidth]{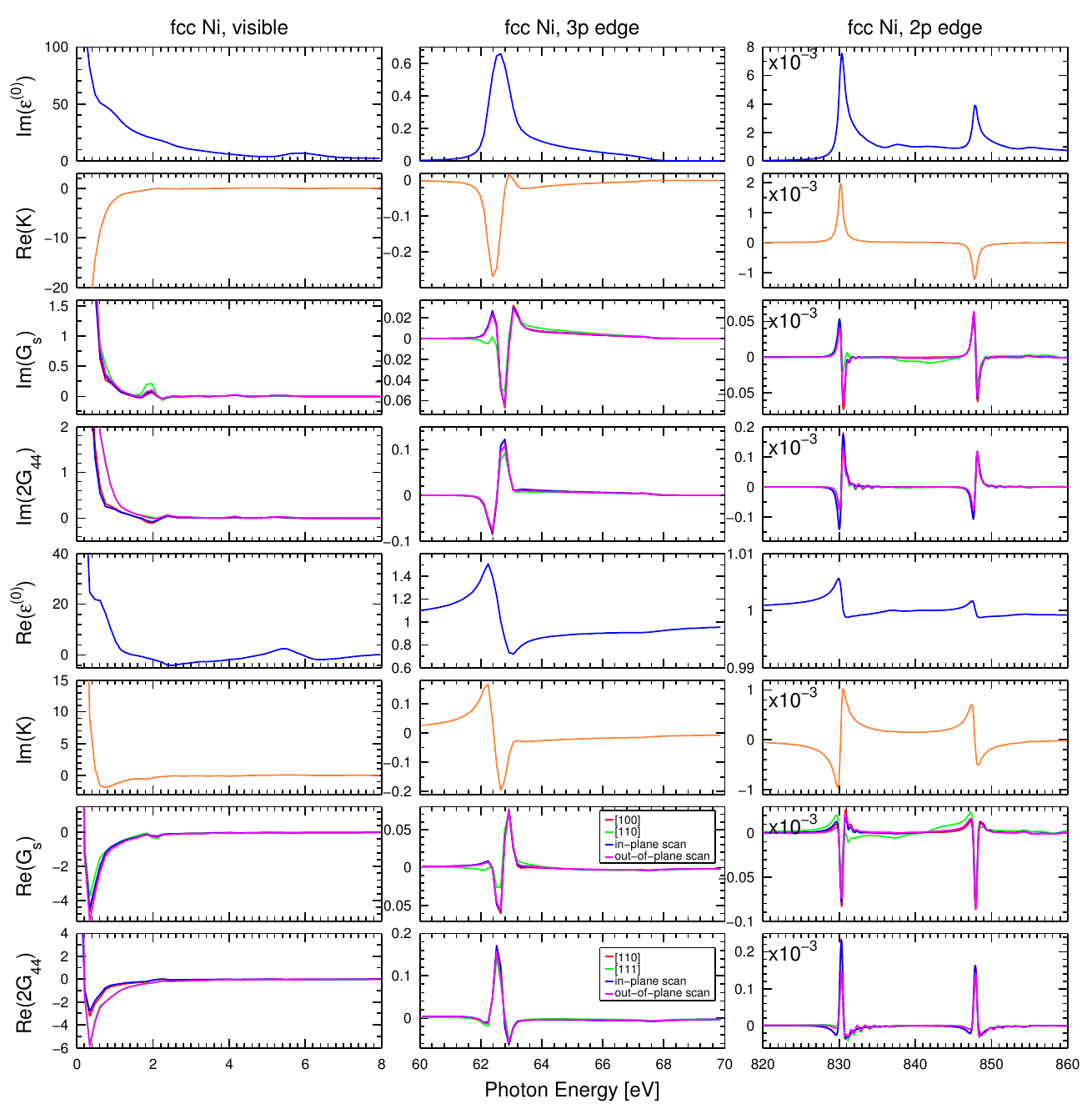}
\caption{\label{f:specNi}%
Principal spectra for fcc Ni. Remaining description is the same as in Fig.~\protect\ref{f:specFe}.
}
\end{figure*}

To determine all principal spectra we first separate linear and quadratic contributions, by extracting off-diagonal linear in magnetization ($(\ep_{ij}-\ep_{ji})/2$, $i\neq j$), off-diagonal quadratic in magnetization ($(\ep_{ij}+\ep_{ji})/2$, $i\neq j$) and diagonal quadratic in magnetization ($(\ep_{ii}-\ep_{jj})/2$, $i\neq j$) contributions from which the principal spectra are determined. The examples of angular dependencies of those contributions are demonstrated in Fig.~\ref{f:epsdiff} for Fe at photon energies 1.3\,eV and 56\,eV.
In Figs.~\ref{f:specFe}--\ref{f:specNi} we show principal spectra of $\ep^{(0)}$, $K$, $G_{s}$, $2G_{44}$ for bcc Fe, fcc Co and fcc Ni. 
As those materials own point O$_h$ symmetry $\Delta\Gamma=0$. The parts of the spectra related with light absorption are shown in first four lines, whereas last four lines are related to light dispersion (determined by Kramers-Kronig relations \cite{rathgen2004}).

Two approaches to determine principal spectra are used.  
(i) We determine principal spectra from ab-initio calculated $\ep_{ij}$ for in-plane and out-of-plane magnetization scans. For both scans, the angular dependences as predicted by symmetry arguments for cubic crystals are presented in Tab.~\ref{t:scans}. Knowing those dependences we can extract all principal spectra. Two pairs of spectra of $G_s$ and $2G_{44}$ are determined, each pair from one scan.
For computational details, see Appendix~\ref{app:comp}.
(ii) We determine principal spectra solely from $\ep_{ij}$ calculated for magnetization pointing in fundamental symmetry directions such as [100], [110] or [111] being sufficient to determine all principal spectra. Tab.~\ref{t:fund} summarizes analytical expressions of principal spectra from known permittivity elements with magnetization at fundamental magnetization directions. Two pairs of spectra are determined, spectra of $G_s$ from [100] and [110] magnetization directions, and spectra of $2G_{44}$ from [110] and  [111] magnetization directions.

Obviously, when calculated angular dependences follow relations predicted by symmetry arguments, and when we neglect contributions of higher order effects, both approaches providing four pairs of spectra should be identical. All those principal spectra are shown in Figures~\ref{f:specFe}-\ref{f:specNi} for $G_s$ and $2G_{44}$. One can see that although all elements and spectral ranges were calculated from equal electronic structure, there is very good agreement between all approaches in case of 2$p$- and 3$p$-edges spectral range. On the other hand, agreement between $G$-spectra in visible range is much smaller (particularly visible for $G_s$ in Fe). This can be also seen on Fig.~\ref{f:epsdiff}, where demonstrated example (Fe at 1.3\,eV) where, in case of both in-plane scan and out-of-plane scans, the values of e.g.\ $\ep_{yy}-\ep_{zz}$ or $\ep_{zz}-\ep_{xx}$ are shifted. 
The discrepancy stems from the numerical precision of the Brillouin zone integration technique used to calculate optical properties for photon energies below about 1--2\,eV. For more details see the Appendix~\ref{app:comp}.

Finally, recall, for isotropic materials $G_{s}=2G_{44}$. However, in 2$p$-edge and 3$p$-edge spectra, the peaks of $2G_{44}$ and $G_{s}$ corresponding to L$_{2,3}$ and  M$_{2,3}$ edges have roughly opposite signs, demonstrating very strong (dominating) anisotropy of quadratic effects in cubic bcc Fe, fcc Co and fcc Ni in 
their respective spectral ranges.
Very strong anisotropy signal in XMLD is demonstrated and analysed by both experimental and theoretical calculations at both 2$p$-edge and 
3$p$-edge in case of the bcc Fe and fcc Co \cite{Nol10,valencia10,tesch2014,leg14,mertins14} and compare the corresponding experimental spectral shapes with 
spectral shape of here predicted G-tensor elements ($G_{s}, G_{44}$) related to the quadratic contributions of magneto-optical effects.
Note also the determined Im$(K)$, element related to the linear magneto-optical effects (middle column of Fig.~\ref{f:specFe}) has the same spectral shape as recently measured T-MOKE spectra of bcc Fe at the 3$p$ edges \cite{leg15}. 

Furthermore, Fig.~\ref{f:specFe} demonstrates (visible particularly for 3$p$-edges) that light absorption $\Im(\ep^{(0)})$ has form of sum of Lorentzian peaks. Then the 
absorption of linear element $K$ is roughly proportional to the first derivation of $\ep^{(0)}$: $\Re(K)\sim \partial \Im(\ep^{(0)})/\partial E$, and quadratic elements 
are approximately proportional to second derivative of $\ep^{(0)}$: $\Im(2G_{44}),\,\Im(G_{s})\sim \partial^2 \Im(\ep^{(0)})/\partial E^2$ \cite{kunes2003}. 
Notice also different spectral features in both $G$-spectra, e.g., in case of Fe at 4.7\,eV (or Ni at 1.9\,eV), there is a strong peak in  $G_{s}$ spectra, but no such a feature in $2G_{44}$ spectra. Similar can be told about feature for Ni at 0.4\,eV, presented only in $2G_{44}$ spectra and absent in $G_{s}$ spectra.

Finally notice that for (001) oriented crystals, (a) the real/imaginary part of $G_{s}$ is proportional to magnetic linear birefringence/dichroism (MLB/MLD), respectively, with $\vec{M}\parallel[100]$.
Namely, it corresponds to difference of light speed/absorption for $E\parallel[100]$ and $E\parallel[010]$.
Similarly, $2G_{44}$ is proportional to MLB/MLD with $\vec{M}\parallel[110]$, i.e.\ $E\parallel[110]$ and $E\parallel[1\overline{1}0]$. Once more let us stress that  
spectral shapes of G-tensors can be directly compared with recorded XMLD spectra w.r.t. various M and E vectors of Refs.\  
\cite{Nol10,valencia10,tesch2014,leg14,mertins14}. This can be seen also seen on expressions of $Gs$ and $2G_{44}$ using Kubo formula (\ref{eq:sigs}--\ref{eq:sig44}) 
below. 
 (b) In case of both $\vec{M}$ and $\vec{E}$ vectors laying in (001)-plane (i.e. in-plane magnetization and normal light incidence), the isotropic part of MLB/MLD, is proportional to the term $(G_{s}+2G_{44})/2$, whereas the anisotropic part of MLB/MLD (i.e.\ varying with crystallographic orientation) is given by $\Delta G=G_{s}-2G_{44}$ \cite{Laan11}.

\section{Analytical quantum-mechanical description}

We express analytically all principal spectra ($\ep^{(0)}$, $K$, $G_{s}$, $2G_{44}$) by quantum mechanical description (i.e.\ by modified Kubo formula) within the electric dipole approximation.
In Cartesian system defined $\hat{x}\parallel[100]$, $\hat{y}\parallel[010]$, $\hat{z}\parallel[001]$ and for magnetization along $M\parallel[001]$ direction, the absorption (dissipative) part is the well-known non-magnetic (isotropic) term $\ep^{(0)}=\ep_{zz}$ is \cite{bennett1965,oppeneer1992,bruno1996} 

\begin{equation}
   \omega \Im(\ep^{(0)}(\omega))=\frac{\pi e^2}{\hbar\omega m^2\Omega}
   \sum_{i,f}[f(E_f)-f(E_i)]
   \left|\langle i|\vec{\pi}\cdot\hat{z}|f\rangle\right|^2 
   \delta(\omega-\omega_{fi})\,.
   \label{eq:sig0}
\end{equation}
where $\langle i|\vec{\pi}\cdot\hat{z}|f\rangle$ is matrix elements for light polarization (electric field) along $\hat{z}$ with
$|i\rangle$, $|f\rangle$ being initial and final electron states, respectively. Kinetic momentum operator
$\vec{\pi}=\vec{p}+\frac{e}{c}\vec{A}+\frac{\hbar}{4mc^2}\vec{\sigma}\times\nabla U$ is sum of canonical momentum $\vec{p}$, vector potential of the magnetic field $\vec{A}$ and spin-orbit coupling between electron potential $U$ and its spin $\sigma$, $f(E_i)$, $f(E_f)$ are occupation functions. The dispersive part of the dielectric functions are determined by Kramers-Kronig relations \cite{rathgen2004}. Note that according to Tab.~\ref{t:scans}, Eq.~(\ref{eq:sig0}) calculates imaginary part of $\ep_{zz}=\ep^{(0)}+G_{11}$, not solely $\ep_{zz}=\ep^{(0)}$. However, $G_{11}$ can not be determined, as length of magnetization vector is constant for ferromagnetic materials. Hence, $G_{11}$ is inseparable part of $\ep^{(0)}$ and we neglect it compared to $\ep^{(0)}$.

When both magnetization and light propagation is along $\hat{z}$ axis, $M\parallel[001]$, then eigenmodes of light propagation are circularly left and circularly right polarizations. Therefore, 
the linear term $K$ is given as a difference of light absorption for circularly-left and circularly-right polarizations as determined by Bruno \textit{et al} \cite{bruno1996}
\begin{equation}
   -\omega \Re(K(\omega))=\frac{\pi e^2}{4\hbar\omega m^2\Omega}
   \sum_{i,f}[f(E_f)-f(E_i)]
   \left( |\langle i|\vec{\pi}\cdot(\hat{x}+i\hat{y})|f\rangle|^2 - |\langle i|\vec{\pi}\cdot(\hat{x}-i\hat{y})|f\rangle|^2 \right)
   \delta(\omega-\omega_{fi})\,.
   \label{eq:sigK}
\end{equation}


In case magnetization is along $M\parallel[100]$ direction (i.e.\ along $\hat{x}$), the symmetry arguments predict $G_{s}=\ep_{xx}-\ep_{yy}$ (Tab.~\ref{t:scans}). By employing Eq.~(\ref{eq:sig0}), $G_s$ can be expressed as
\begin{equation}
   \omega \Im(G_s)=\frac{\pi e^2}{\hbar\omega m^2\Omega}
   \sum_{i,f}[f(E_f)-f(E_i)]
   \left( |\langle i|\vec{\pi}\cdot\hat{x}|f\rangle|^2 - |\langle i|\vec{\pi}\cdot\hat{y}|f\rangle|^2 \right)
   \delta(\omega-\omega_{fi})\,.
   \label{eq:sigs}
\end{equation}
This equation corresponds to the case when 
light propagation is along $\hat{z}$ axis and $M\parallel\hat{x}$, then eigenmodes of light propagation are linearly polarized waves parallel and perpendicular to magnetization direction. Hence, the quadratic term $G_s$ is given by difference of absorptions for those two light polarizations.

Similarly, in Cartesian system defined $\hat{x}'\parallel[110]$, $\hat{y}'\parallel[\overline{1}10]$, $\hat{z}'\parallel[001]$ and for magnetization $M\parallel[110]$, the element $2G_{44}=\ep_{x'x'}-\ep_{y'y'}$ 
is expressed formally in the same way as $G_s$ \cite{hamrlova2013}. Transforming back to Cartesian system $\hat{x}\parallel[100]$, $\hat{y}\parallel[010]$, $\hat{z}\parallel[001]$, 
quadratic term $2G_{44}$ is expressed by difference of absorption of linear light polarization along [110] and $[1\overline{1}0]$ directions when magnetization is along $[110]\parallel \hat{x}+\hat{y}$
\begin{equation}
   \omega\Im(2G_{44})=\frac{\pi e^2}{2\hbar\omega m^2\Omega}
   \sum_{i,f}[f(E_f)-f(E_i)]
   \left( |\langle i|\vec{\pi}\cdot(\hat{x}+\hat{y})|f\rangle|^2 - |\langle i|\vec{\pi}\cdot(-\hat{x}+\hat{y})|f\rangle|^2 \right)
   \delta(\omega-\omega_{fi})\,.
   \label{eq:sig44}
\end{equation}
Finally note, that here expressed $G_s$ and $2G_{44}$ corresponds to principal spectra calculated from [100] and [110] magnetization directions, as denoted in Tab.~\ref{t:fund} and Figs.~\ref{f:specFe}--\ref{f:specNi}

\section{Relation between permittivity tensor elements and experimental techniques}

The permittivity tensor phenomenologically describes optical properties of matter. In this Section, we provide a brief relation between elements of permittivity tensor and measured effects in common magnetooptical techniques.

From experimental point of view, the measured quantities are straightforwardly related with the reflectivity matrix, usually expressed in Jones formalism \cite{fujiwara-book}. Then, measured optical quantities, such as Kerr effect, magnetic circular dichroism (MCD), MLD, reflectivity, magnetoreflectivity, ellipsometry etc.\ are directly related to this reflection matrix. In case of multilayer structure, the relation between reflection matrix and permittivities of constituent layers are routinely described by Yeh \cite{yeh1980} or Berreman \cite{berreman1972} formalisms, based on propagation of light eigenmodes in the layer, where eigenmodes' intensities are determined by boundary conditions on the interfaces (continuity of transversal component of electric and magnetic fields through the interfaces). As optical anisotropies are usually week, the perturbation optical approach has been developed as well \cite{bertrand2001,censor2002}, handling optical anisotropy as a perturbation from optical isotropic response. 

Analytical solution of optical response of multilayer structure with generally optically anisotropic layers is very complicated. Hence, analytical solution is mostly expressed for two simple structures: (i) reflection on interface of bulk material. However, even in this simple case the analytical expression of reflection matrix from generally optically anisotropic media is still very complicated \cite{hunt1966, mansuripur1990, zak1991,visnovsky-book}. (ii) ultrathin approximation, where thickness of anisotropic layer is assumed to be much thinner than wavelength of light in the material \cite{hamrle2003, visnovsky1995, visnovsky-book}. In this approximation, optical response can be expressed analytically rather simply. 

Although different multilayer structures provides different relations between constituent permittivity tensor and reflection matrix, some relations can be generalized. In following we assume one of the following simplifications:
\begin{itemize}
\item[(a)] many magnetooptical technique (e.g.\ MOKE, transverse MOKE (TMOKE) or XMCD) are selectively sensitive to off-diagonal permittivity elements of the FM layer. In this case, to describe effect measured by the setup, we neglect difference between the diagonal permittivity elements, and assume all diagonal permittivity elements to be equal. 
\item[(b)]
on contrary, MLD employs change in reflection/transmission of linearly polarized light. Hence, when handling intensities of those reflected or transmitted linearly polarized light, we can neglect influence of the off-diagonal permittivity elements and assume them vanished.
\end{itemize}

In case of (a) approximation, the reflection matrix element can be written in general form (neglecting higher orders in $\ep_{ij}$, $i\neq j$) as
\begin{align}
r_{sp}&=a \ep_{yx} + b \ep_{zx}
\\
r_{ps}&=-a \ep_{xy} + b \ep_{xz}
\\
r_{pp}&=r_{pp}^{(0)} + c (\ep_{yz}-\ep_{zy})
\\
r_{ss}&=r_{ss}^{(0)}
\end{align}
where $a$, $b$, $c$, $r_{ss}^{(0)}$ and $r_{pp}^{(0)}$ are optical terms given by details of optical and geometrical properties of the multilayer structure, being independent on the off-diagonal permittivity of the FM layer  $\ep_{ij}$, $i\neq j$. Note that for zero incidence angle ($\varphi=0$), $r_{ss}^{(0)}=-r_{pp}^{(0)}$

As experimentally measured magnetooptical quantities are directly related to reflection matrix, we demonstrate relation between off-diagonal permittivity elements and few magnetooptic techniques:

\subsubsection{Magneto-optic Kerr effect (MOKE)}

The complex MOKE angle $\Phi=\theta+i\epsilon$ ($\theta$ is Kerr rotation, $\epsilon$ is Kerr ellipticity) can be expressed as 
\begin{align}
 \Phi_s & =-\frac{r_{ps}}{r_{ss}} = A_s \ep_{yx} + B_s \ep_{zx}
 \\
 \Phi_p & = \frac{r_{sp}}{r_{pp}} = -A_p \ep_{xy} + B_p \ep_{xz} 
\end{align}
where $A_{s/p}$, $B_{s/p}$ are again optical terms depending on all other optical properties of the constituent layers \cite{ham07quad}. Note that for $\varphi=0$, $A_s=A_p$ and $B_s=B_p=0$, and for small incidence angle $B_s\approx B_p$.

\subsubsection{Transverse MOKE (TMOKE)}

TMOKE is change of p-reflectivity due to off-diagonal elements is described by term $\Delta r_{pp} = c(\ep_{yz}-\ep_{zy})$, being usually induced by transverse magnetization.
Then, TMOKE writes 
\begin{equation}
  \mathrm{TMOKE}=\frac{\Delta r_{pp}}{r_{pp}} = T (\ep_{yz}-\ep_{zy})
\end{equation}
where again $T$ is optical term depending on optical details of the multilayer.

\subsubsection{Magnetic circular dichroism (MCD)}

MCD (and obviously also XMCD) is difference of light absorption for circularly left and right polarized light. To be consistent with previous part, we express MCD in reflection,  MCD=$R_\mathrm{right}-R_\mathrm{left}$, where $R_\mathrm{right}$, $R_\mathrm{left}$ is reflection for circularly right, left polarized light, respectively:
\begin{equation}
 \mathrm{MCD}=R_\mathrm{right}-R_\mathrm{left}=2\Im (r_{ss} r_{sp}^* - r_{pp} r_{ps}^*)
\end{equation}
where $^*$ denotes complex conjugation. In any case, the relations between off-diagonal permittivity elements and MCD in transmission are formally identical. Assuming small incidence angle ($\varphi\approx0$), and hence $r_{ss}\approx-r_{pp}$, we got
\begin{equation}
  \mathrm{MCD}=2\Im(r_{ss} (r_{sp}^*+r_{ps}^*)) = A_\mathrm{MCD} (\ep_{yx} -\ep_{xy}) + B_\mathrm{MCD} (\ep_{zx}+\ep_{xz})
\end{equation}
where again $A_\mathrm{MCD}$ and $B_\mathrm{MCD}$ are optical terms depending on optical details of the structure.

\subsubsection{Magnetic linear dichroism (MLD)}
In case of MLD (or XMLD), the approximation (b) is employed, assuming off-diagonal term vanished. Then, off-diagonal reflectivity coefficients vanishes as well ($r_{sp}=r_{ps}=0$) and diagonal reflectivities can be written as
\begin{align}
  r_{ss}&=r_{ss}^{(0)} + u \Delta\ep_{xx} 
  \\
  r_{pp}&=r_{pp}^{(0)} + v \Delta\ep_{yy} +w \Delta\ep_{zz}
\end{align}
where $u$, $v$ and $w$ are again optical terms and $\Delta\ep_{xx}$, $\Delta\ep_{yy}$ $\Delta\ep_{zz}$ are differences of diagonal permittivity elements from from $\ep^{(0)}$ (i.e.\ differences between anisotropic and isotropic permittivities).
Note in case $\varphi=0$  term $w$ vanish and $u=-v$. Also, in case of metals, term $w$ can be neglected as well even for non-zero incidence angle due to large refractive index of metals. MLD is defined as a difference of reflectivity (or transmittivity) between incident $s$ and $p$ polarized light. Further assuming $\varphi=0$, MLD writes
\begin{equation}
  \mathrm{MLD}=R_{ss} -R_{pp} = |r_{ss}|^2 - |r_{pp}|^2 = U (\Delta\ep_{xx}-\Delta\ep_{yy})
\end{equation}
where $U$ is again an optical term.

\section{Conclusion}

In conclusion, we provide principal spectra, {\textit i.e.} decomposition of isotropic, linear and quadratic contributions in magnetization of magnetooptic effects up to the second order in magnetization. The values of magnetooptic permittivity elements are determined as weighted sum of principal spectra, where weights are determined solely by magnetization direction. Although in X-ray range, this spectroscopy is already established, it was not introduced for visible range. 
In case of ferromagnetic cubic crystals owning point symmetry, all magneto-transport up to second-order effects are described by spectra of four complex principal spectra, denoted $\varepsilon_0$, $K$, $G_{s}$ and $2G_{44}$, as predicted by symmetry arguments. This approach expresses second-order magnetooptic properties as bulk material property, independent on sample crystallographic orientation, layer thickness, structure details etc. The principal spectra were expressed from ab-initio calculations for bcc Fe, fcc Co and fcc Ni. To extract principal spectra from ab-initio, we have used two methods (i) by comparing ab-initio calculated angular dependence of the permittivity tensor, with dependences predicted by symmetry arguments and (ii) using permittivity elements calculated for fundamental symmetry magnetization directions. 
We also express principal spectra analytically using modified Kubo formula. 
Although the spectroscopy presented here is for cubic crystals owning $O_h$ point symmetry, it can also be simply modified for materials with different crystal symmetries, higher orders in magnetizations or even different magneto-transport effects. 
We expect that experimental and theoretical determination of principal spectra will be routinely used in magnetooptic spectroscopy of ferromagnetic materials.

\begin{acknowledgments}
Financial support by Czech Grant Agency (13-30397S), by IT4Innovations Centre of Excellence project (CZ.1.05/1.1.00/02.0070), funded by the European Regional Development Fund and the national budget of the Czech Republic via the Research and Development for Innovations Operational Programme, as well as Czech Ministry of Education via the project Large Research, Development and Innovations Infrastructures (LM2011033) and  Student Grant Competitions of VSB-TU Ostrava (SP2015/61, SP2015/71 and SP2016/182) as well as helpful discussions with Peter M. Oppeneer and Karel V\'yborn\'y are well acknowledged.   
\end{acknowledgments}

\appendix
\section{Computational details}
\label{app:comp}
The angular dependence of $\ep_{ij}$ on magnetization direction with respect to the crystal axes is determined by ab-initio calculations.
The electronic structure is determined by full potential linearised augmented plane wave method (WIEN2k code \cite{Wien2k}), where the exchange energy is based on local spin density approximation \cite{LSDA}. The spin-orbit coupling was included in a second-variation scheme \cite{Kunes01}. The following parameters were employed: the energy cutoff, given as the product of the muffin-tin radius and the maximum reciprocal space vector  was $R_\mathrm{MT}K_\mathrm{max}=8.5$, the largest reciprocal vector in the charge Fourier expansion, $G_\mathrm{max}$, was set to 14 Ry$^{1/2}$, and the maximum value of partial waves inside the muffin-tin spheres, $l_\mathrm{max}=10$. The Brillouin zone sampling was done on grid of 27$\times$27$\times$27 $k$-points, whereas for the optical transition matrix elements \cite{Amb06} a finer grid of 81$\times$81$\times$81 was used. The optical calculations are based on Kubo formula \cite{oppeneer1992}, with omitted Drude term, hence we do not present dc transport. At present, also 
core-hole effects are omitted. The lattice constants of studied materials are fixed being $a=2.870$\AA(bcc Fe), $a=3.425$\AA(fcc Co) and $a=3.420$\AA(fcc Ni) \cite{CK}.

The numerical precision shows to be limited for the beginning of the optical energy range below about 1--2\,eV,{\it i.e.} the states in vicinity of the Fermi level.
There may be two reasons for it:
(1) the calculated transition probabilities are calculated for each pair of energy levels and for each $k$-point of the mesh of the reciprocal space. The integration of the Kubo formula is 
implemented by Bl\"ochl-enhanced tetrahedron integration \cite{bloechl1994}, which determines mean value of a given property (e.g.\ transition probability) inside each tetrahedron. This mean value is calculated by weighted sum of  a given property at each corner of the tetrahedron, where the weights for each stationary state (i.e.\ for each energy level) are calculated from eigenenergy of each k-point and value of the Fermi energy. This works well, when either initial or final tetrahedron are fully above or below Fermi level. However, in case when both initial and final states are within tetrahedrons cutting Fermi level, this approach may not work properly. This may explain, why dependence of permittivity elements on magnetization direction for photon energy below 1--2\,eV are rather different from expected symmetry dependence compared to those calculated for higher photon energies.
(2) In Bl\"ochl-enhanced tetrahedron integration, the irreducible $k$-points are selected to fulfil that the symmetry operation copies those irreducible $k$-points to all $k$ points inside the unit cell of the reciprocal space. Obviously, in the case of high symmetry structure, such as cubic crystal, the position of $k$-points in the unit cell follows most of symmetry operations. However, the integration employs tetrahedrons, which does not follow so many symmetry operations. For example, in case of cubic crystal, the rotation by 90$^\circ$ copy $k$-points to themselves, but the tetrahedrons are not copied to themselves. Therefore, the numerical error of the integration from different parts of the unit cell does not follow structural symmetry, which may increase error described in the previous paragraph.

%

\end{document}